\documentclass[showpacs, twocolumn]{revtex4-1}
\usepackage{graphicx}
\usepackage{amsmath}
\usepackage{appendix}

\begin{document}

\title{Quantum correlations in dipolar droplets: 
Time-dependent Hartree-Fock-Bogoliubov theory}

\author{Abdel\^{a}ali Boudjem\^{a}a and Nadia Guebli}
\affiliation{ Department of Physics,  Faculty of Exact Sciences and Informatics, Hassiba Benbouali University of Chlef P.O. Box 78, 02000, Ouled Fares, Chlef, Algeria.}
\email {a.boudjemaa@univ-chlef.dz} 


\begin{abstract}

We investigate the effects of quantum correlations on dipolar quantum droplets.
To this end, we derive self-consistent time-dependent Hartree-Fock-Bogoliubov equations that fairly describe the dynamics of the order parameter, 
the normal, and anomalous quantum correlations of the droplet. 
We analyze the density profiles, the critical number of  particles, the condensate depletion, and the pair correlation function.
Our predictions are compared with very recent experimental and Quantum Monte-Carlo simulations results and excellent agreement is found.

\end{abstract}


\maketitle

\section{Introduction}

Successful realization of quantum droplets in dipolar Bose-Einstein condensates (BEC) \cite{Pfau1, Pfau2,Chom}
has opened an avenue to exploit a broad range of exciting physical phenomena of interest such as supersolid states \cite{Mondu, Pfau3, Chom2} and rotonic stripe phases \cite{Pfau4}.
Quantum droplets are stabilized by quantum fluctuations and require a minimum number of particles to be stable. 
This exquisite stabilization mechanism originates from the balance of the attractive mean-field energy and repulsive beyond mean-field effects. 

Recently, dipolar self-bound droplets have been the object of intense theoretical investigations 
\cite{Pfau5, Wach,Saito, Bess2, Wach1, Bess3,Boudj1, Bess1, Kui, Mac, Mish, Bail, Pfau6, Boudj2, Boudj3, Pfau7}. 
Many of the results coming from these studies, such as ground-state properties, lifetimes, and excitation frequencies of the droplet, 
are based on the zero-temperature generalized Gross-Pitaevskii (GGP) equation, which includes the Lee-Huang-Yang (LHY) corrections 
\cite{Pfau2, Wach, Saito, Bess2, Wach1, Bess3,Chom}. 
The main disadvantage of the GGP equation is that it fails to predict the critical atom number compared to experiments \cite {Chom,Chom2} 
and to describe effects of quantum correlations on the droplets \cite{Pfau7}.  
These correlations manifest in significant deviations of critical number of particles, Bose condensate depletion, and pair correlation function of 
self-bound droplets.

In this work, we develop a versatile theoretical model describing the experiment realized by B\"ottcher {\it et al}. \cite{Pfau7}. 
Our analysis is based on the self-consistent time-dependent Hartree-Fock-Bogoliubov (GTDHFB) equations 
where the order parameter is coupled to the normal and anomalous correlations \cite{Boudj2, Boudj4, Boudj5,Boudj6}. 
Previous models neglected these, difficult to compute, terms. 
The TDHFB equations are based on the Balian-V\'en\'eroni variational principle  \cite{BV}.
This latter requires that both the state of the system and the observable of interest vary in their own variational space.
The main difference between the TDHFB approach and the other variational treatments is that, in our variational theory, 
we do not minimize only the expectation values of a single operator such as the free energy in the variational HF and HFB approximations but 
we minimize an action in addition to a Gaussian variational ansatz.
In spite of the Gaussian character of the variational ansatz, the approximations obtained in this way go beyond the usual mean-field theory
and include correlations between particles \cite{Boudj4, BV,Cecile}. These latter permit us to extract in a useful way the pair correlation function.
The TDHFB equations extend naturally the GGP equation \cite{Pfau1, Pfau2,Chom, Pfau5, Wach,Bess2, Wach1, Bess3, Bess1} and the HFB-Popov  theory \cite {Boudj1} 
as well as the HFB approximation \cite{Ayb} since they include selfconsistently the dynamics of the quantum correlations.
They inherently contain beyond-mean-field corrections due the presence of quantum depletion and anomalous density in contrast to the GGP model.
An important feature of our theory is that it remains valid for any number of particles in contrast to the Quantum Monte-Carlo (QMC) method used in \cite{Pfau7} 
which is limited only for  small number of particles.

We show that the quantum fluctuations induced by interactions can be exactly determined using the TDHFB formalism, 
reproducing the seminal LHY results \cite{LHY, lime, Boudj7,Boudj8}.
By providing a numerical implementation of the full TDHFB equations, we analyze the behavior of the density profiles 
for various atomic numbers and interaction strengths.
Our results reveal that the interactions and the number of particles may modify the shape of different densities.
For relatively strong interactions, the anomalous density exhibits a special structure near the center.
To the best of our knowledge, the anomalous correlations have never been examined before for the self bound droplet.
Furthermore, we calculate the critical number of particles, the condensate depletion, and the pair correlation function. 
It is found that our theory captures genuine quantum correlations effects predicted from experiment and QMC simulations \cite{Pfau7}.


The rest of this paper is structured as follows. 
In Sec.\ref{Theory}, we present a description of the main features of the current TDHFB theory of dipolar droplets.
Section \ref{results} is devoted to the numerical results of the full TDHFB equations. 
We calculate the density profiles, the critical number of  particles, the condensate depletion, and the pair correlation function. 
The findings are compared with available experimental and QMC data.
Finally, our conclusions are summarized in Sec.\ref{Conc}.


\section{Time-dependent Hartree-Fock-Bogoliubov equations theory}  \label{Theory}

We consider a dipolar BEC with dipoles are oriented along the $z$-direction and  in a cylindrically symmetric confinement, 
$U({\bf r})=\frac{1}{2}m \omega_r^2 (r^2+\lambda^2 z^2)$,
where $r^2=x^2+y^2$, and $\lambda=\omega_z/\omega_r$ is the ratio between the trapping frequencies in the axial and radial directions.
The behavior of dipolar BECs including the effect of the LHY quantum corrections and normal and anomalous correlations is described by
the nonlocal TDHFB equations which can be written in the compact form as \cite{Boudj2, Boudj4, Boudj5,Boudj6}: 
\begin{subequations}\label{TDHFB}
\begin{align}
&i\hbar \dot{\Phi} = \bigg [h^{sp}+ \int d{\bf r'} V({\bf r}-{\bf r'}) n ({\bf r'}) + \delta \mu_{\text{LHY}} \bigg]\Phi,  \label{TDHFB1}\\
&i\hbar \frac{d \rho}{d t} =-2\left[\rho, \frac{ \partial {\cal E}}{\partial \rho}\right], \label{TDHFB3} 
\end{align}
\end{subequations}
where  $h^{sp}= -(\hbar^2/ 2m) \Delta + U-\mu$ is the single particle Hamiltonian, and $\mu$ is the chemical potential.
The density matrix $\rho ({\bf r, r'})$ contains both normal $\tilde n ({\bf r, r'})=\langle \hat {\bar\psi}^\dagger ({\bf r}) \hat {\bar\psi} ({\bf r'}) \rangle$
and anomalous $\tilde m ({\bf r, r'})= \langle \hat {\bar\psi} ({\bf r}) \hat {\bar\psi} ({\bf r'}) \rangle$ components with 
$\hat {\bar \psi} ({\bf r})=\hat\psi ({\bf r})- \Phi ({\bf r})$ being the noncondensed part of the field operators $\hat\psi (\bf r)$. It is defined as \cite{Boudj2, Boudj4, Boudj5,Boudj6}: 
\begin{equation} \label{DMO}
\rho ({\bf r},{\bf r'})=\begin{pmatrix} 
\tilde n ({\bf r, r'}) &&& \tilde m ({\bf r, r'})\\
\tilde m^* ({\bf r, r'})&&& \tilde n^* ({\bf r, r'})+\delta ({\bf r}-{\bf r'})
\end{pmatrix}. 
\end{equation}
In Eq.(\ref{DMO}), the normal and anomalous correlation functions represent the dipole exchange interaction between the condensed and noncondensed atoms. 
In the local limit they reduce, respectively  to the noncondensed  $\tilde n ({\bf r})$ and anomalous $\tilde m ({\bf r})$ densities.
It is worth stressing that the density operator (\ref{DMO}) satisfies the positivity condition: if $\rho \geq 1$ at $t=0$, then this holds for all times.
Such positivity preservation property implies that
i) $\tilde n ({\bf r, r'}) \geq 0$, and $\tilde m ({\bf r},{\bf r'})= \tilde m ({\bf r'},{\bf r})$ (symmetric).
ii) The equality $\rho (\rho+1)=0$ which yields the relation (\ref{Inv1}) holds true in the sense of quadratic forms.
iii) The conservation of the von Neumann entropy.
Furthermore, the hermiticity of the density matrix implies that $\tilde n ({\bf r, r'})$ is real.
The total density is given by $n ({\bf r})=n_c({\bf r})+\tilde n ({\bf r})$ with $n_c({\bf r})=|\Phi({\bf r})|^2$ being the condensed density.
The energy of the system reads
\begin{align}  \label{enrgyfunc}
&{\cal E} (\Phi, \tilde n, \tilde m)= \int d {\bf r} h^{sp} ({\bf r})  [\tilde n ({\bf r},{\bf r})+\Phi({\bf r}) \Phi^*({\bf r}) ] \\
&+\int d {\bf r} d {\bf r'} V ({\bf r}-{\bf r'}) |\Phi ({\bf r})|^2 |\Phi ({\bf r'})|^2\nonumber \\
&+\frac{1}{2}\int d {\bf r} d {\bf r'} V ({\bf r}-{\bf r'}) [\tilde m^* ({\bf r},{\bf r'})\tilde m ({\bf r},{\bf r'}) \nonumber\\
&+ \tilde n ({\bf r},{\bf r'}) \tilde n ({\bf r'},{\bf r}) +\tilde n ({\bf r'},{\bf r'}) \tilde n ({\bf r'},{\bf r'})] \nonumber\\
&+\frac{1}{2}\int d {\bf r} d {\bf r'} V ({\bf r}-{\bf r'}) [\tilde n ({\bf r},{\bf r'})\Phi({\bf r}) \Phi^*({\bf r'}) \nonumber\\
&+\tilde n ({\bf r'},{\bf r})\Phi^*({\bf r}) \Phi({\bf r'}) +\tilde n ({\bf r'},{\bf r'})\Phi({\bf r}) \Phi({\bf r})+ \tilde n ({\bf r},{\bf r})\Phi({\bf r'}) \Phi({\bf r'}) ] \nonumber\\
&+\frac{1}{2}\int d {\bf r} d {\bf r'} V ({\bf r}-{\bf r'}) [\tilde m^* ({\bf r},{\bf r'})\Phi({\bf r}) \Phi({\bf r'}) 
+ \tilde m ({\bf r},{\bf r'})\Phi^*({\bf r}) \Phi^*({\bf r'}) ]. \nonumber
\end{align} 
The two-body interactions potential reads $V({\bf r})=g\delta({\bf r})+[C_{dd}(1-3\cos^2\theta) ]/ (4\pi r^3)$,
where $g=4\pi \hbar^2 a/m$ corresponds to the short-range part of the interaction that is parametrized by the $s$-wave scattering length $a$. 
The DDI coupling constant $C_{dd}$ is characterized by the dipole-dipole distance  $r_*=m C_{dd}/4\pi \hbar^2$ determined by the magnetic moment
($r_*=129 a_0$ for ${}^{162}$Dy, and $r_*=131 a_0$ for  ${}^{164}$Dy \cite {Pfau7} with $a_0$ being the Bohr radius),
 and $\theta$ is the angle between ${\bf r}$ and the polarization axis.

The right-hand side of Eq.(\ref{TDHFB3}) can be evaluated as 
$$\left[\rho, \frac{d{\cal E}}{d\rho}\right]_{ij}= \sum_{\ell=1}^2\left(\rho_{i \ell} \left(\frac{\partial {\cal E}}{\partial \rho}\right)_{ \ell j}- 
\left(\frac{\partial {\cal E}}{\partial\rho}\right)_{i \ell} \rho_{\ell j} \right),\;\; i,j=1,2.$$
Explicit equations of motion for the matrix $\rho$ elements, i.e.($\tilde n$ and $\tilde m$), can be obtained easily by employing the usual convention:
$(\partial /\partial\rho)_{ij}=\partial/\partial\rho_{ji}$. 

At zero temperature,  the TDHFB equations (\ref{TDHFB}) are coupled with another useful equation linking the normal and anomalous densities \cite{Boudj4,Cecile}
\begin{align}  \label{Inv1}
\tilde n ({\bf r},{\bf r'})&= \sqrt{\frac{1}{4} \delta ({\bf r}-{\bf r'}) + \int d {\bf r''} \tilde m^*({\bf r},{\bf r''})  m({\bf r''},{\bf r'}) } \\
&-\frac{1}{2} \delta ({\bf r}-{\bf r'}). \nonumber
\end{align}
Evidently, this equation which steems from the conservation of the Von Neumann entropy \cite{Cecile}, 
shows that the anomalous correlations is larger than the normal correlation even at zero temperature signaling the 
importance of $\tilde m$ in ultracold Bose gases. 
Equation (\ref{Inv1}) permits us to eliminate the variable $\tilde n ({\bf r})$ and thus, the TDHFB equations reduce to the coupled equations for the order parameter 
and the anomalous correlation.
Equivalently, by applying the local density approximation (LDA), one can also express the relation (\ref{Inv1}) using the definitions:
$\tilde n ({\bf r}) =\sum_{\bf k} v_{\bf k}^2 ({\bf r})$, and $\tilde m ({\bf r})=-\sum_{\bf k} u_{\bf k} ({\bf r}) v_{\bf k} ({\bf r})$, 
where $ u_{\bf k} ({\bf r}) v_{\bf k} ({\bf r})=[\sqrt{\varepsilon_k ({\bf r}) /E_k}\pm\sqrt{E_k/\varepsilon_k ({\bf r})} ]/2$ are the Bogoliubov functions with $E_k= \hbar^2k^2/2m$ 
and $\varepsilon_k ({\bf r})$ is the local excitations energy \cite{Boudj6}. 
Therefore, solving  Eq.(\ref{Inv1}) locally recovers the GGP used in the literature.

Importantly, in our formalism the relevant  LHY corrections to the equation of state are obtained self-consistently \cite{Boudj2}  
\begin{align} \label{LHY}
 \delta \mu_{\text{LHY}} ({\bf r}) \Phi({\bf r})&=\int d{\bf r'} V ({\bf r}-{\bf r'}) \bigg[\tilde n ({\bf r},{\bf r'})\Phi({\bf r'}) \\
&+\tilde m ({\bf r},{\bf r'})\Phi^*({\bf r'})\bigg].  \nonumber
\end{align}
The LDA evaluation of this term together with Eq.(\ref{Inv1}), after dimensional regularization of the anomalous density \cite{Boudj2,Boudj8}, gives
$g_{\text{LHY}} n_c^{3/2}$ \cite{Boudj7,Boudj8}, where $g_{\text{LHY}} \simeq (32/3) g\sqrt{a^3/\pi} (1+3\epsilon^2_{dd}/2)$ 
is the strength of the LHY quantum corrections with $\epsilon_{dd}=r_*/a$ \cite{Pfau2, Bess2, Boudj1, Boudj7, Boudj8}.
The dimensional regularization of the anomalous density used here \cite{Boudj2, Boudj8} is equivalent to the Born approximation
for the interaction potential of Ref.\cite{ldz} and valid at finite temperatures  \cite{Boudj2, Boudj8}.
The contribution of the anomalous density term in the TDHFB equations (\ref{TDHFB}) is of second order in the effective
interaction potential.
Remarkably,  the condensed  density $n_c$ which constitutes our corrections, appears as a key parameter instead of the total density $n$ found in \cite{lime}.

Equations (\ref{TDHFB})-(\ref{Inv1}) are self-consistent and non-divergent equations.
They take into account the dynamics of the normal and anomalous correlations without involving any assumptions. 
Importantly, Eqs. (\ref{TDHFB3}) provide a sophisticate description of the measured physical quantities such as the critical number of particles \cite{Pfau7}, 
since they allow us to produce a lower negative energy, ${\cal E}$, than that predicted by the GGP equation (see below). 
The usefulness of the TDHFB theory appears in its accuracy at making quantitative predictions in both BEC (low density) \cite{Boudj4, Boudj5,Boudj6} 
and droplet (hight density) regimes at zero and finite temperatures \cite{Boudj2}. 
TDHFB theories similar to our own have been derived  for BEC using different approaches \cite {Bijl,Griffin, Giorg, Chern,Corn,Yuk,Prouk}. 

\section {Results} \label{results}

The density profiles of a static droplet can be found by solving iteratively and selfconsistently  the renormalized time-independent version of Eqs.(\ref{TDHFB})-(\ref{Inv1}).
The numerical simulation was performed using the split-step Fourier transform and the convolution method to remove the singularity of the DDI at the origin \cite{Boudj3,Ron}. 
Once the stability is checked we proceed with the cylindrical symmetry in order to compute the normal and anomalous exchange contributions 
(for more details see  Ref.\cite{Corm}).

\begin{figure}
\includegraphics[scale=0.45]{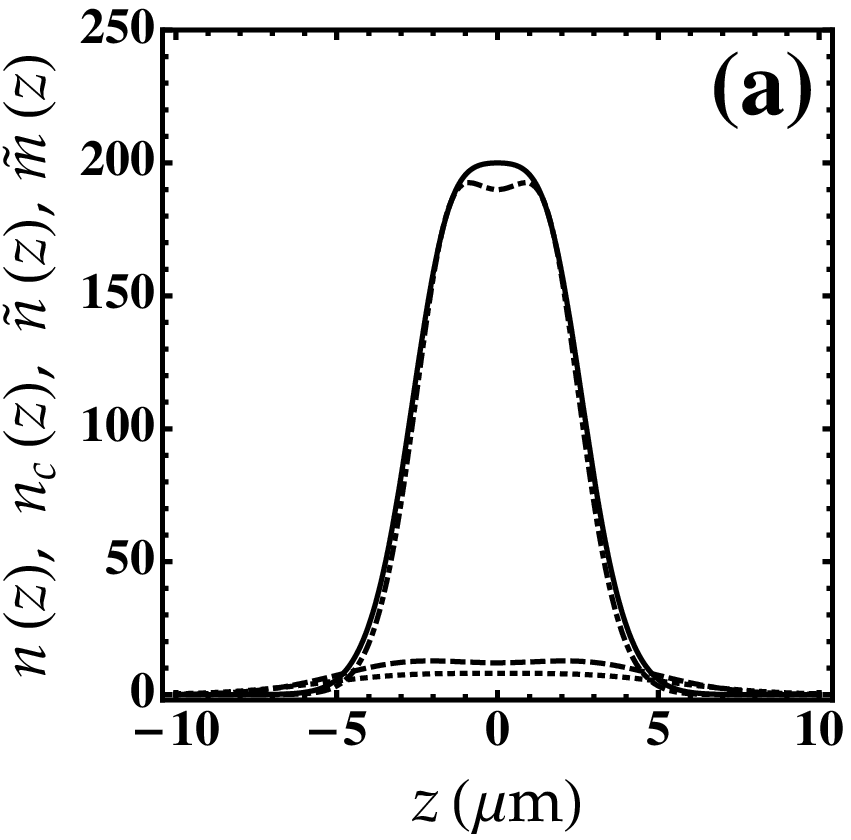}
\includegraphics[scale=0.45]{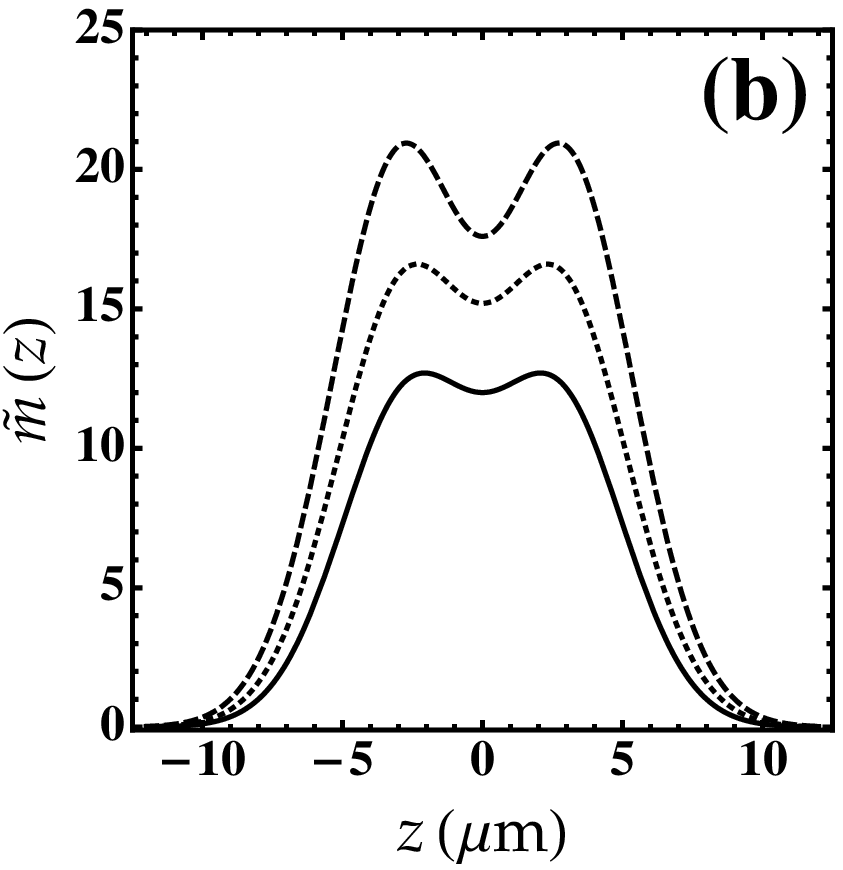}
\includegraphics[scale=0.45]{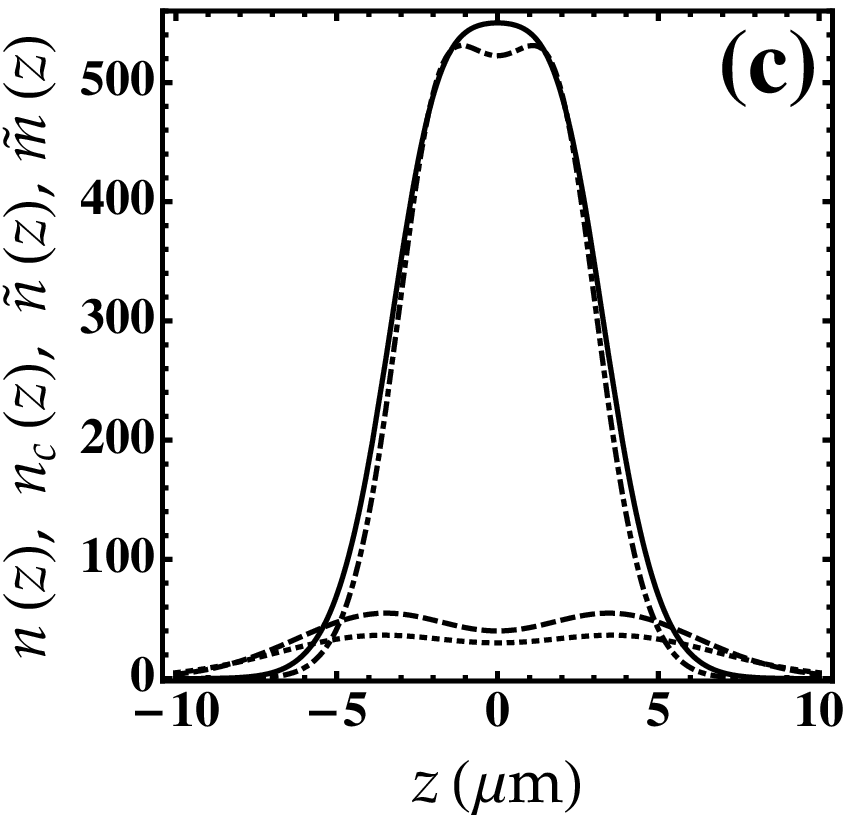}
\includegraphics[scale=0.45]{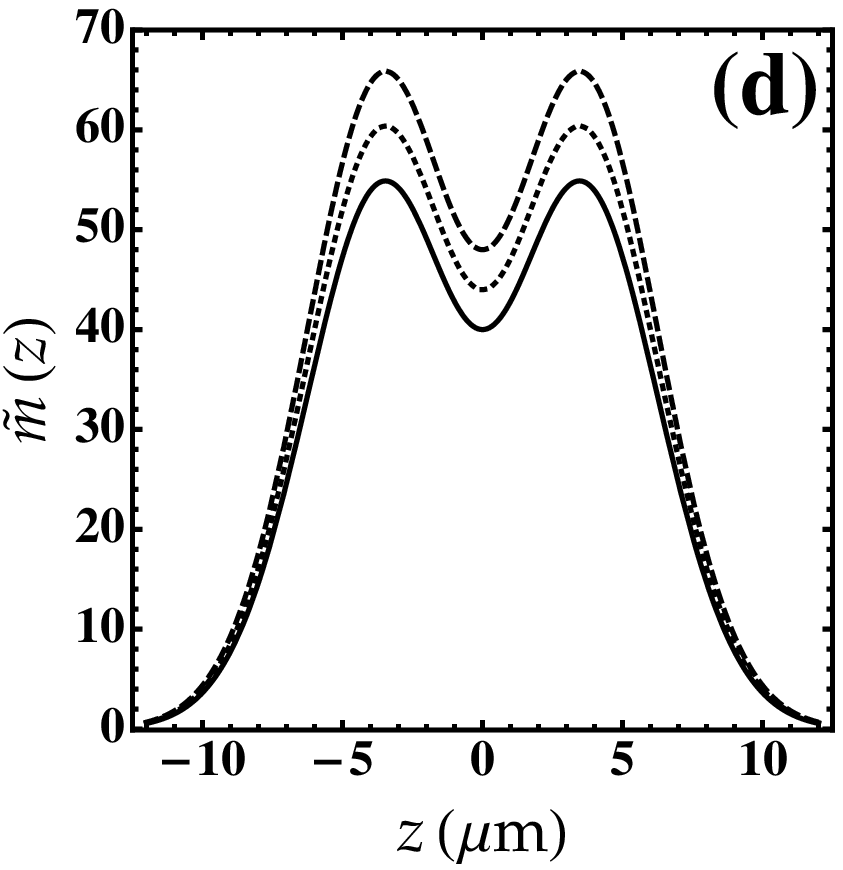}
\caption {(a) Total density $n$ (solid lines), condensed density $n_c$ (dotdashed lines), noncondensed density $\tilde n$ (dotted lines) and anomalous density $\tilde m$ (dashed lines)  
along the axial direction of the droplet for $N=2\times 10^3$ atoms and a scattering length of $a= 60 a_0$.
(b) Anomalous density for different values of $a$.  Solid lines: $a= 60 a_0$, dotted lines: $a= 90 a_0$, and dashed lines: $a= 120 a_0$.
(c)-(d) The same as panels (a) and (b) but for $N=10^4$. Here interactions can be modulated thanks to Feshbach resonance.
The noncondensed and the anomalous densities have been amplified ten times for clarity.}
\label{DP}
\end{figure}

As can be seen in Fig.\ref{DP} (a), the condensate, the noncondensate population and the anomalous correlations are intertwinned.
The size of the noncondensed and anomalous densities is larger than that of the condensate. 
Similar behavior holds true in the two-fluid model of liquid helium \cite{FG}. 
The condensed density is enhanced in the low density surface region (edges) of the self-bound droplet in the sense that the LHY corrections become 
smaller and the main competition is between the mean-field term and the kinetic energy. 
In this low density surface region where the system is purely Bose condensed, the order parameter can be described using the standard TDHFB equations.
We observe also that the total density $n$ is comparable to the condensed density since the local condensate depletion is very small at zero temperature (curves with dotted lines). 
One should stress that the depletion is quite small for quantum droplets even in the bulk in stark contrast with the superfluid ${}^4$He droplets (see e.g.\cite{Lew,Griffin1}).
In this latter case, the surface region can be regarded also as an inhomogeneous dilute Bose gas but with $n_c \approx 0.1 n$ \cite{Lew,Griffin1}.

Figure \ref{DP} (b) shows that the anomalous correlations are larger than the normal correlation of the Bogoliubov particles $\tilde m$
even at zero temperature as is foreseen in Eq.(\ref{Inv1}). 
Surprisingly, the dip in the central density of $\tilde m$ is suppressed due to the dilutness of the droplet which is in contrast to the trapped BEC case \cite{Boudj4, Hut}.
Such a dip appears near the center of the anomalous density only for relatively strong interaction ($a=120 a_0$) where
$\tilde m$ increases itself with the scattering length.

Figures \ref{DP} (c) and \ref{DP} (d)  depict that the three densities are increasing with the number of particles, $N$.

\begin{figure}
\includegraphics[scale=0.8]{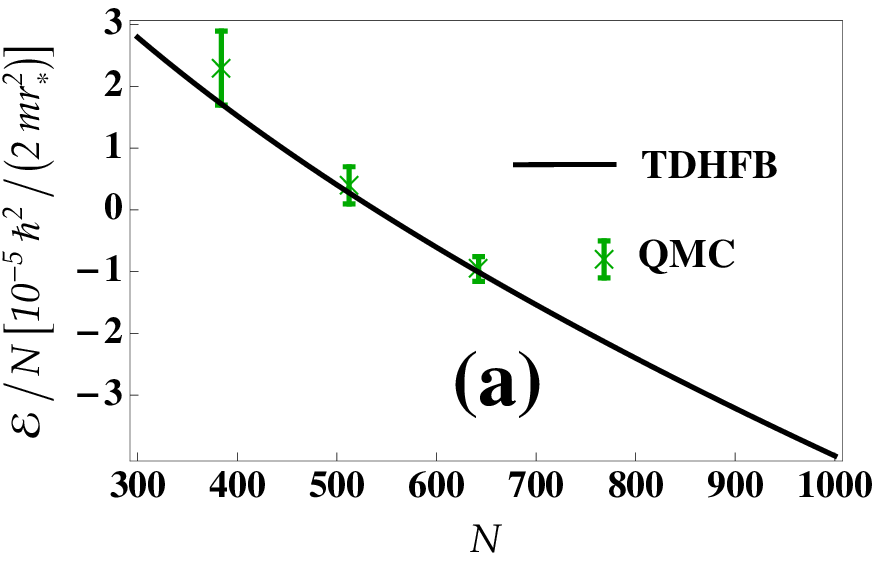}
\includegraphics[scale=0.8]{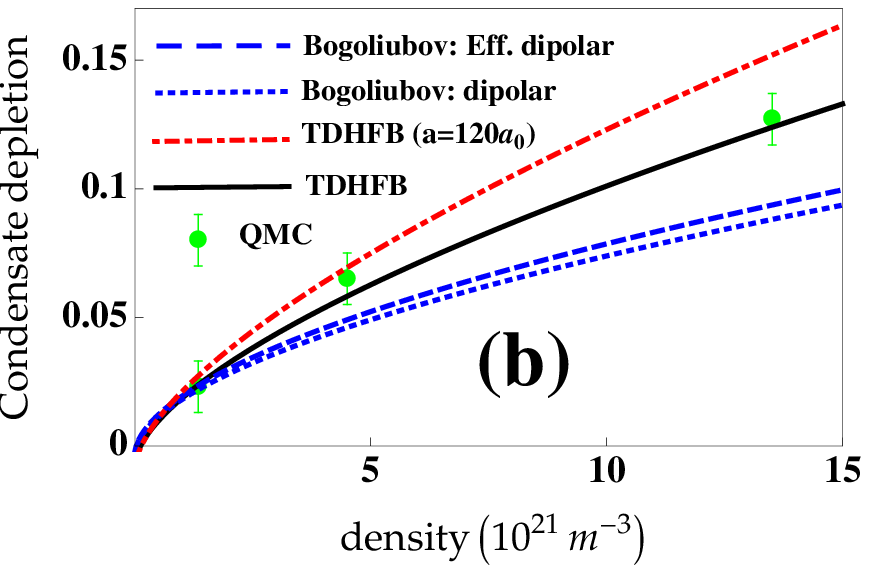}
\caption {(Color online) (a) Energy per particle from Eq.(\ref{enrgyfunc}) for the $s$-scattering length $a= 60a_0$.
(b) Condensate depletion for $a= 60 a_0$ (black solid line) and $a= 120 a_0$ (red-dot-dashed line).}
\label{Depl}
\end{figure}

We now turn our attention to evaluating the condensate depletion and the pair correlation function and compare our results with 
QMC simulation points \cite{Pfau7} and the Bogoliubov theory \cite{Pfau7}.
To this end, we consider a homogeneous bulk ($U=0$) with a density of $n=5.88 \times 10^{21}$ m$^{-3}$, corresponding to the central density of a saturated quantum droplet. 
In this regime, Eqs.(\ref{TDHFB}) do not admit uniform solutions with constant $\Phi$, $\tilde n$ and $\tilde m$ but instead they
admit solutions with a negative energy which is in perfect agreement with QMC \cite {Pfau7}  as is shown in Fig.\ref{Depl} (a) indicating the formation of a stable self-bound droplet. 

In Fig.\ref{Depl} (b) we plot our predictions for the condensate depletion as a function of the density.
We find that the TDHFB predictions improve the Bogoliubov treatment. 
This improvement which is lost in the GGP and the Bogoliubov methods, brings our theory in excellent agreement with QMC simulations of \cite{Pfau7}.
In the same figure, we observe that the presence of dipolar interactions may enhance the condensate depletion which is in best agreement with  
previous theoretical results \cite{Boudj1, lime, Boudj7,Boudj8}.
Strong atomic interactions, which lead to considerable effects of the anomalous correlation,
can eject atoms from the droplet to the noncondensate component, yielding large quantum depletion (see the red-dot-dashed line). 
This is in accordance with recent experimental results obtained for weakly interacting BEC \cite{Lops}.

The second-order (pair) correlation function, $g^{(2)} ({\bf r},{\bf r'})=\langle \hat\psi^\dagger({\bf r})  \hat\psi^\dagger({\bf r'})\hat\psi ({\bf r'})\hat\psi({\bf r}) \rangle$,
is a fundamental quantity to characterize the coherence of the self-bound droplet state.
In our formalism it couples to normal and anomalous correlations as:
\begin{align}\label {2Corr2}
g^{(2)} ({\bf s})=n_c^2+2n_c\int_0^{\infty}  \frac{d {\bf k} }{(2\pi)^3}\bigg (\tilde{n}+\tilde{m} \bigg) e^{i {\bf k}. \bf s},
\end{align}
where ${\bf s}=|{\bf r}-{\bf r'}|$.
This expression clearly shows that the pair correlation function depends on the anomalous density.
The absence of $\tilde m$ may affect the long-range behavior of $g^{(2)}(s)$.
This latter can be obtained by inserting the normal and anomalous correlations found from Eqs.(\ref{TDHFB})-(\ref{Inv1}) into Eq.(\ref{2Corr2}).
We then carry out a quantitative comparison to the recent experimental and QMC simulations results of \cite{Pfau7}.

\begin{figure}
\includegraphics[scale=0.8]{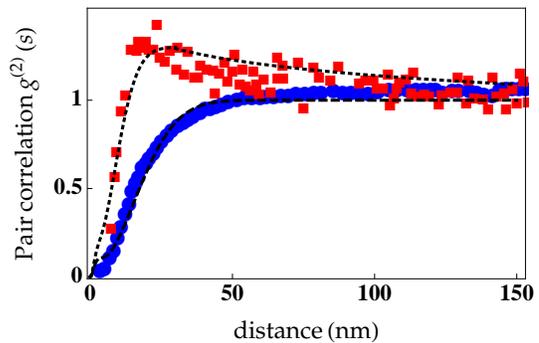}
\caption {(Color online) Pair correlation function $g^{(2)}(s)$ at a density of $n=5.88 \times 10^{21}$ m$^{-3}$, 
corresponding to the central density of a saturated quantum droplet at $a= 60 a_0$.
Red squares and blue circles indicate directions perpendicular and parallel to the polarization axis, respectively, correspond to the  QMC data \cite{Pfau7}.
Black dashed (parallel) and dotted (perpendicular) lines correspond  to our TDHFB predictions.}
\label{CorrF}
\end{figure}

Figure \ref{CorrF} captures the behavior of the pair correlation function $g^{(2)}$  at an equilibrium density of $n=5.88 \times 10^{21}$ m$^{-3}$.
Because of the anisotropy of the DDI, $g^{(2)}$ splits into two different directions parallel and perpendicular to the dipole orientation. 
In the perpendicular direction, the pair correlation function vanishes at short distances and then increases monotonically similarly to ordinary weakly interacting Bose gases. 
Whereas in the parallel direction, $g^{(2)}$ grows significantly and develops a peak due to the interplay of the DDI and quantum correlations. 
For even larger quantum correlations, $g^{(2)}$ may exhibit a steeper peak at small distances leading to reduce the BEC coherence 
results in a new quantum phase transition.
Impacts of DDI on the pair correlation function have been highlighted in our recent work \cite{Boudj7}.
It is clearly visible that the curves of the TDHFB approach and QMC \cite{Pfau7} agree with each other in both directions.
The good accordance with the QMC simulation reflects that the long-range behavior of $g^{(2)}$ is accurately described by our theory.
Experimentally, the pair correlation function can be measured using either a Bragg diffraction interferometer \cite{Cac} or four-wave mixing
of the collision of two BECs \cite{Perin}.

\begin{figure}
\includegraphics[scale=0.8]{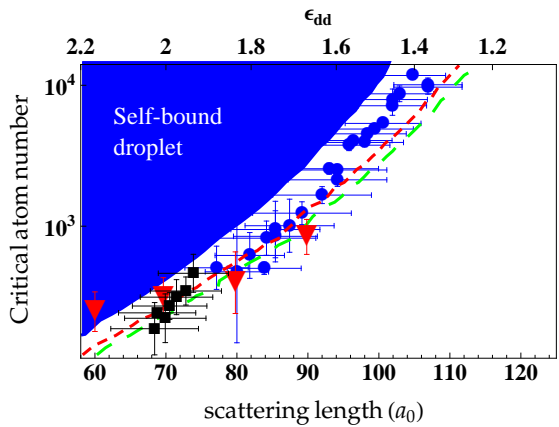}
\caption {Critical atom number of a self-bound dipolar quantum
droplet for ${}^{162}$Dy (blue points) and ${}^{164}$Dy \cite{Pfau5} (black squares). Blue line shows the numerical solutions of the GGP equation \cite{Pfau7}.
The red dashed line shows the corresponding boundary as expected from an increased effective dipolar length due to a finite collision energy 100 nK  \cite{ldz}. 
The red triangles show the results obtained by QMC simulations \cite{Pfau7}.
Green dashed line corresponds to our TDHFB predictions.}
\label{CNP}
\end{figure}

Let us now focus on the critical number of particles $N_{cr}$ of the self-bound droplet. 
We extract the number of particles using the method outlined in \cite{Pfau7} and choose $U= 0$. 
We then compare our findings with available experimental data, GGP equation and QMC results. 
Figure \ref{CNP} shows that the critical atom number obtained from the TDHFB  theory is lower than that of the GGP equation 
even with an effective enhancement of the dipolar length.
This downward shift of $N_{cr}$ which is in good agreement with the experimental results and QMC simulations of \cite{Pfau7}, is due to the effects of qunatum correlations.
We see also that $N_{cr}$ is rising with the scattering length but still harmonious with the experimental error bars.
The critical atom number for self-bound droplets can be measured by applying a magnetic field gradient along the $z$-direction after the preparation of the BEC \cite{Pfau7}.


\section{Conclusions} \label{Conc}

In this paper, we investigated the effects of quantum correlations on the dipolar droplet state using the TDHFB theory
in which the LHY term is derived without involving any subsidiary assumptions.
The condensate, noncondensate and anomalous density profiles of the droplet have been deeply analyzed. 
Our results revealed that the condensed density is enhanced in low density surface region of the droplet.
The anomalous density is increased with interactions and number of atoms and its shape is modified due to crucial role of the quantum correlations.
We demonstrated that our TDHFB theory is able to produce excellent predictions for the condensate depletion,
and the pair correlation function that have been measured experimentally and with QMC \cite{Pfau7}.  
We elucidated the crucial role played by normal and anomalous quantum correlations on critical number of particles and found
that this latter is lowered owing to the intriguing effects of the quantum correlations. 
Results showed that the TDHFB predictions, and QMC simlations are in very good agreement with each other.
Our results pave the way for investigations on effects of quantum correlations on the behavior of supersolid and striped phases in dilute dipolar droplets.


\section*{Acknowledgments}
We would like to thank Tilman Pfau and Fabian B\"ottcher for providing us with experimental data.
We are grateful to Ferran Mazzanti for the QMC data and for useful comments and suggestions. 
We acknowledge support from the Algerian Ministry of Higher Education and Scientific Research under Research Grant No. PRFU-B00L02UN020120190001.

\newpage

\end{document}